\documentclass[aps,a4paper,twocolumn,superscriptaddress]{revtex4}
\usepackage{graphicx}

\newcommand{\beq}{\begin{equation}}
\newcommand{\eeq}{\end{equation}}
\newcommand{\beqa}{\begin{eqnarray}}
\newcommand{\eeqa}{\end{eqnarray}}

\renewcommand{\a}{\alpha}
\newcommand{\abs}[1]{\vert#1\vert}
\newcommand{\at}[1]{\langle#1\rangle}
\renewcommand{\c}{{({\rm c})}}
\newcommand{\dd}{{\rm d}}
\newcommand{\daut}[1]{\frac{\dd#1}{\dd\tau}}
\newcommand{\dpar}{\partial}
\newcommand{\e}{{\rm e}}
\newcommand{\frad}[2]{{\displaystyle{\displaystyle{#1}\over\displaystyle{#2}}}}
\newcommand{\lam}{\lambda}
\newcommand{\mean}[1]{\langle#1\rangle}
\newcommand{\kave}{{\mean{k}}}
\renewcommand{\r}{{({\rm r})}}
\newcommand{\s}{\sigma}
\newcommand{\w}{\widetilde}
\newcommand{\z}{\zeta}
\newcommand{\F}{{_{2}F_{1}}}
\newcommand{\Lam}{\Lambda}
\newcommand{\T}{{\bf T}}

\begin{document}

\title{Universality in survivor distributions:\\
Characterising the winners of competitive dynamics}

\author{J.~M. Luck}\email{jean-marc.luck@cea.fr}
\affiliation{Institut de Physique Th\'eorique, Universit\'e Paris-Saclay, CEA
and CNRS,
91191 Gif-sur-Yvette, France}

\author{A. Mehta}\email{anita@bose.res.in}
\affiliation{S.~N. Bose National Centre for Basic Sciences, Block JD, Sector 3,
Salt Lake, Calcutta 700098, India}

\begin{abstract}
We investigate the survivor distributions of a spatially extended model of competitive dynamics
in different geometries.
The model consists of a deterministic dynamical system of individual agents
at specified nodes, which might or might not survive the predatory dynamics:
all stochasticity is brought in by the initial state.
Every such initial state leads to a unique
and extended pattern of survivors and non-survivors, which is known as an
attractor of the dynamics.
We show that the number of such attractors grows exponentially with
system size, so that
their exact characterisation is limited to only very small systems.
Given this, we construct an
analytical approach based on inhomogeneous mean-field theory to calculate
survival probabilities for arbitrary networks.
This powerful (albeit approximate)
approach shows how universality arises in survivor distributions via a key concept --
the {\it dynamical fugacity}.
Remarkably, in the large-mass limit,
the survival probability of a node becomes independent of network geometry, and assumes
a simple form which depends only on its mass and degree.
\end{abstract}

\maketitle

\section{Introduction}
\label{intro}

The fate of agents in any situation where death is a possibility
attracts enormous interest: here, death does not have to be literal,
but could also refer to bankruptcies in a financial context or oblivion
in the context of ideas.
This panorama of situations is usually captured by agent-based models
including predator-prey models,
the best known being the Lotka-Volterra model~\cite{murray,beco,sfpr} or indeed,
winner-takes-all models, which embody the survival of the largest, or the fittest.
The model we study here belongs to the second category; the most massive agents grow at
the cost of their smaller neighbors, which eventually disappear.
Motivated by the physics of interacting black holes in brane-world cosmology~\cite{bh1,bh2},
its behavior in mean-field and on different lattices was investigated at length in~\cite{us},
followed by simulations on more complex geometries~\cite{c1,c2}.
The essence of the model
in its original context~\cite{bh1,bh2} involved the competition between black holes of different masses, in the presence of a universal dissipative `fluid'.
The model also turned out to have a `rich-get-richer' interpretation in the context
of economics, where it was related to the survival dynamics
of competing traders in a marketplace in the presence of taxation (dissipation)~\cite{c3,c4}.

One of the most important questions to be asked of such models concerns the
distribution of survivors, i.e., those agents who survive the predatory dynamics.
In spatially extended models, the presence of multiple interactions
makes this a difficult question to answer in full generality.
The seemingly universal distributions of survivor patterns
put forward in~\cite{c1} motivated us to ask: can one provide a theoretical framework
for the appearance of some universal features in survivor distributions?
The answer is yes, as we will demonstrate in this paper.

We first introduce a much simpler version of the model investigated
earlier~\cite{us,c1,c3,c2}, which, however, preserves features such as the exponential
multiplicity of attractors (Section~\ref{model}) essential to its complexity.
We characterize exactly the attractors reached by the dynamics
on chains and rings of increasing sizes (Section~\ref{exact}),
an exercise which illustrates how a very intrinsic complexity
makes such computations rapidly impossible.
This leads us to formulate an approximate analytical treatment
of the problem on random graphs and networks (Section~\ref{ana}),
based on inhomogeneous mean-field theory,
which yields rather accurate predictions for
the survival probability of a node, given its degree and/or its initial mass.

\section{The model}
\label{model}

The degrees of freedom of the present model consist of a
time-dependent positive mass $y_i(t)$ at each node $i$ of a graph.
These masses are subject to the following first-order dynamics:
\beq
\frac{\dd y_i}{\dd t}=\left(1-g\sum_{j(i)}y_j\right)y_i.
\label{dydt}
\eeq
Each node $i$ is symmetrically (i.e., non-directionally) coupled to its neighbors,
which are all nodes $j$ connected to~$i$ by a bond;
$g$ is a positive coupling constant.

The dynamics are deterministic, so that all stochasticity
comes from the distribution of initial masses $y_i(0)$.
This model is a much simpler version of the one investigated in~\cite{us}
which was inspired by black-hole physics~\cite{bh1,bh2}.
In particular the explicit time dependence and
initial big-bang singularity of that model are here dispensed with;
only a simple nonlinearity, quadratic in the masses, is retained.
Despite these simplifications, the present model keeps the most interesting features
of its predecessor, such as those to do with its multiplicity of attractors.

The coupling constant $g$ can be scaled out by means of a linear rescaling of the masses:
\beq
z_i(t)=gy_i(t).
\eeq
The new dynamical variables indeed obey
\beq
\frac{\dd z_i}{\dd t}=\left(1-\sum_{j(i)}z_j\right)z_i.
\eeq

As a matter of fact, the model has a deeper dynamical symmetry.
Setting
\beq
y_i(t)=a_i(\tau)\,\e^t,
\eeq
where $\tau$ is a global proper time, so that $\dd\tau=g\,\e^t\,\dd t$, i.e.,
\beq
\tau=g(\e^t-1),
\eeq
the dynamical equations~(\ref{dydt}) can be recast as
\beq
\daut{a_i}=-\left(\sum_{j(i)}a_j\right)a_i.
\label{dadtau}
\eeq
Remarkably, the dynamics so defined are entirely para\-me\-ter-free.
Quadratic differential systems such as the above have attracted much attention
in the mathematical literature, such as in discussions of Hilbert's 16th
problem (see e.g.~\cite{jen,li}).

Equations~(\ref{dadtau}) can be formally integrated as
\beq
a_i(\tau)=a_i(0)\,\exp\left(-\int_0^\tau\sum_{j(i)}a_j(\tau')\,\dd\tau'\right).
\label{expl}
\eeq
The amplitudes $a_i(\tau)$ are therefore decreasing functions of $\tau$.
For each node $i$, either of two things might happen:

\begin{itemize}

\item Node $i$ survives asymptotically.
This occurs when the integral in~(\ref{expl}) converges in the $\tau\to\infty$ limit.
The amplitude $a_i(\tau)$ reaches a non-zero limit $a_i(\infty)$,
so that the mass $y_i(t)$ grows exponentially~as
\beq
y_i(t)\approx a_i(\infty)\e^t.
\label{yexp}
\eeq

\item Node $i$ does not survive asymptotically.
This occurs when the integral in~(\ref{expl}) diverges in the $\tau\to\infty$ limit.
This divergence is generically linear,
so that the amplitude $a_i(\tau)$ falls off to zero exponentially fast in $\tau$,
while the mass $y_i(t)$ falls off as a double exponential in time~$t$.

\end{itemize}

The dynamics therefore drive the system to a non-trivial attractor,
i.e., an extended pattern of survivors and non-survivors.
This attractor depends on the whole initial mass profile
(although it is independent of the overall mass scale).
The formula~(\ref{expl}) generically implies the following local constraints:

\begin{enumerate}

\item each survivor is isolated (all its neighbors are non-survivors),

\item each non-survivor has at least one survivor among its neighbors.

\end{enumerate}

Conversely, every pattern obeying the above constraints
is realized as an attractor of the dynamics, for some domain of initial data.
This situation is therefore similar
to that met in a variety of statistical-mechanical models
ranging from glasses to systems with kinetic constraints.
Attractors play the role of metastable states,
which have been given various names,
such as valleys, pure states, quasi-states or inherent
structures~\cite{tap,ks,sw,kw,fv}.
In all these situations the number $M$ of metastable states
grows exponentially with system size $N$ as
\beq
M\sim\e^{N\Sigma},
\label{sdef}
\eeq
where $\Sigma$ is the configurational entropy or complexity.
This quantity is not known exactly in general,
except in the one-dimensional case where it can be determined by means of
a transfer-matrix approach (see Appendix~\ref{app}).
It is relevant to mention the {\it Edwards ensemble} here, which is constructed
by assigning a thermodynamical significance to the configurational entropy~\cite{edw}.
According to the Edwards hypothesis, all the attractors of a given ensemble
(e.g.~at fixed survivor density) are equally probable.
This hypothesis holds generically for mean-field models,
while it is weakly violated for finite-dimensional
systems~\cite{fv,dl,bm,pb,gds,glrev}.

\section{Exact results for small systems}
\label{exact}

In this section we consider the model on small one-dimensional graphs,
i.e., closed rings and open chains of~$N$ nodes.
In one dimension,~(\ref{dydt}) and~(\ref{dadtau}) read
\beqa
\frac{\dd y_n}{\dd t}&=&\left(1-g(y_{n-1}+y_{n+1})\right)y_n,\\
\daut{a_n}&=&-(a_{n-1}+a_{n+1})a_n,
\label{dadtau1}
\eeqa
for $n=1,\dots,N$, with appropriate boundary conditions:
periodic ($a_0=a_N$, $a_{N+1}=a_1$) for rings
and Dirichlet ($a_0=a_{N+1}=0$) for chains.

Equations~(\ref{dadtau1}) are very reminiscent of those defining
the integrable Volterra chain.
The coupling term involves the sum $a_{n-1}+a_{n+1}$ in the present model,
whereas it involves the difference $a_{n-1}-a_{n+1}$ in the Volterra system.
The appearance of a difference is however essential for integrability~\cite{rus1,rus2}.
The present model is therefore not integrable, even in one dimension.

Our goal is to characterize the attractor reached by the dynamics
on small systems of increasing sizes,
as a function of the initial mass profile.
This task soon becomes intractable, except on very small systems,
due to the intrinsic complexity of the model.
The numbers $M_N^\r$ and $M_N^\c$ of these attractors
on rings and chains of~$N$ nodes
are given in Table~\ref{numbers} of Appendix~\ref{app}.
These numbers grow exponentially fast with $N$,
according to~(\ref{sdef}), with $\Sigma$ given by~(\ref{sres}).

\subsubsection*{Ring with $N=2$}

The system consists of two nodes connected by two bonds.
Both attractors consist of a single survivor.
The dynamical equations~(\ref{dadtau1}) read
\beq
\daut{a_1}=\daut{a_2}=-2a_1a_2.
\label{dadc2}
\eeq
The difference $D=a_1-a_2$ is a conserved quantity.
If $a_1(0)>a_2(0)$, the attractor is $\at{1}$ (meaning that only node 1 survives)
and its final amplitude is
\beq
a_1(\infty)=D=a_1(0)-a_2(0),
\eeq
and vice versa.
The survivor is always the node with the larger initial mass.

In the borderline case of equal initial masses,
the integrals in~(\ref{expl}) are marginally (logarithmically) divergent.
Both masses saturate to the universal limit $y_1(\infty)=y_2(\infty)=1/(2g)$,
irrespective of their initial value.

\subsubsection*{Chain with $N=2$}

The two nodes are now connected by a single bond.
The dynamical equations are identical to~(\ref{dadc2}), up to an overall factor of two.
Here too, the more massive node is the survivor.

\subsubsection*{Ring with $N=3$}

The attractors again consist of a single survivor.
Although there is no obviously conserved quantity,
the attractor can be predicted by noticing that
\beq
\daut{}(a_2-a_1)=-(a_2-a_1)a_3.
\eeq
The sign of any difference $a_i-a_j$ is therefore conserved by the dynamics.
In other words, the order of the masses is conserved.
In particular the survivor is the node with the largest initial mass.

\subsubsection*{Chain with $N=3$}

The central node 2 plays a special role,
so that the two attractors are $\at{2}$ and $\at{13}$.
There are two conserved quantities, $D=a_1-a_2+a_3$ and $R=a_1/a_3$.
If $D>0$, the attractor is $\at{13}$ and the final amplitudes read
\beq
\frac{a_1(\infty)}{a_1(0)}=\frac{a_3(\infty)}{a_3(0)}
=\frac{D}{a_1(0)+a_3(0)}.
\eeq
If $D<0$, the attractor is $\at{2}$ and $a_2(\infty)=\abs{D}$.

\subsubsection*{Ring with $N=4$}

The two attractors are the `diameters' $\at{13}$ and $\at{24}$.
The alternating sum $D=a_1-a_2+a_3-a_4$ is a conserved quantity.
If $D>0$, the attractor is $\at{13}$ and $a_1(\infty)+a_3(\infty)=D$.
If $D<0$, the attractor is $\at{24}$ and $a_2(\infty)+a_4(\infty)=\abs{D}$.
The attractor is therefore always the diameter with the larger total initial mass.
The individual asymptotic amplitudes cannot however be determined in general.

\subsubsection*{Chain with $N=4$}

The three attractors are $\at{13}$, $\at{14}$ and $\at{24}$.
The alternating sum $D=a_1-a_2+a_3-a_4$ is a conserved quantity.
This is the first case where the attractor cannot be predicted analytically in general.

\begin{figure}[!ht]
\begin{center}
\includegraphics[angle=-90,width=.7\linewidth]{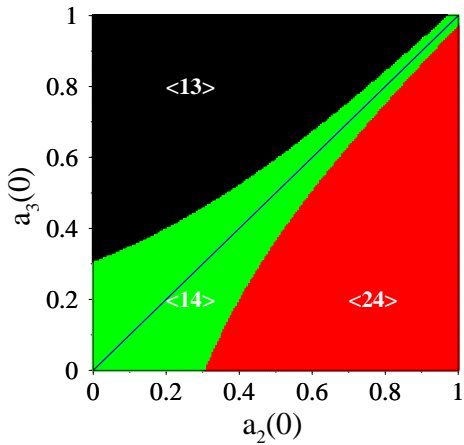}
\caption{\small
(Color online)
Attractor on the chain with $N=4$ in the $a_2(0)$--$a_3(0)$ plane,
for fixed $a_1(0)=a_4(0)=0.3$.}
\label{o4}
\end{center}
\end{figure}

Figure~\ref{o4} shows the attractors reached as a function of $a_2(0)$ and $a_3(0)$,
for fixed $a_1(0)=a_4(0)=0.3$.
It is clear that $\at{13}$ can only be reached for $D>0$, i.e., above the diagonal,
whereas $\at{24}$ can only be reached for $D<0$, i.e., below the diagonal.
The intermediate pattern $\at{14}$ is observed in a central region near the diagonal.
The form of this region can be predicted to some extent.
On the horizontal axis,
the transition from $\at{14}$ to $\at{24}$ takes place for $a_2(0)=a_1(0)=0.3$.
Similarly, on the vertical axis,
the transition from $\at{14}$ to $\at{13}$ takes place for $a_3(0)=a_4(0)=0.3$.
The central region where $\at{14}$ is the attractor shrinks rapidly
with increasing distance from the origin.
This can be explained by considering the dynamics on the diagonal,
i.e., in the symmetric situation where $a_1(0)=a_4(0)$ and $a_2(0)=a_3(0)$.
These symmetries are preserved by the reduced dynamics
\beq
\daut{a_1}=-a_1a_2,\quad
\daut{a_2}=-(a_1+a_2)a_2.
\eeq
The reduced attractor is $\at{1}$, the full attractor is $\at{14}$,
and $D$ vanishes identically.
The reduced dynamics have another conserved quantity, $C=a_1\exp(-a_2/a_1)$.
The asymptotic amplitude $a_1(\infty)=C$
becomes exponentially small as $a_2(0)$ increases.
The width of the central green region is expected to follow the same scaling law,
i.e., to become exponentially narrow with distance from the origin,
in agreement with our observation.

\subsubsection*{Ring with $N=5$}

There are five attractors consisting of two survivors,
obtained from each other by rotation:
$\at{13}$, $\at{24}$, $\at{35}$, $\at{14}$ and $\at{25}$.
There is no obviously conserved quantity,
and the attractor cannot be predicted analytically in general.

\begin{figure}[!ht]
\begin{center}
\includegraphics[angle=-90,width=.7\linewidth]{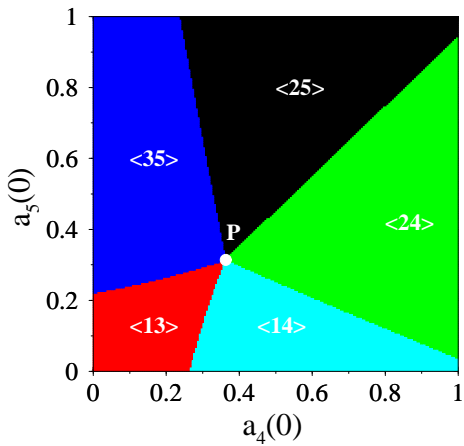}
\caption{\small
(Color online)
Attractor on the ring with $N=5$ in the $a_4(0)$--$a_5(0)$ plane,
for fixed $a_1(0)=0.5$, $a_2(0)=0.7$ and $a_3(0)=0.6$.
The five attractors meet at point P.}
\label{r5}
\end{center}
\end{figure}

Figure~\ref{r5} shows the attractors as a function of $a_4(0)$ and $a_5(0)$,
for fixed $a_1(0)=0.5$, $a_2(0)=0.7$ and $a_3(0)=0.6$.
The five attractors meet at point P $(a_4(0)=0.363094,\;a_5(0)=0.313748)$.
If launched at~P, the system is driven to the unique symmetric
solution where all masses converge to the universal limit $1/(2g)$.
A linear stability analysis around the latter solution
reveals that its stable manifold is three-dimensional,
in agreement with the observation
that its intersection with the plane of the figure is the single point~P.

This is the first case which manifests one of the
most interesting features of the model, that of `winning against the odds'~\cite{c1,c3,c2}.
Numerical simulations with two structureless distributions of initial masses,
uniform (uni) and exponential (exp), yield the following observations:

\begin{itemize}

\item
The probability that the node with largest initial mass
is a survivor is 0.849 (uni) or 0.937 (exp);

\item
The probability that the attractor corresponds to the largest initial mass sum
among the possible attractors is 0.791 (uni) or 0.891 (exp);

\item
The probability that the node with smallest initial mass
is a survivor is 0.018 (uni) or 0.019 (exp).

\end{itemize}

This small but non-zero probability for the smallest initial mass to survive is
the beginning of the complexity associated with the phenomenon
of winning against the odds.
It happens essentially because more distant nodes can destroy massive intermediaries
between themselves and the small masses concerned,
thereby letting the latter survive `against the odds'.
For larger sizes, the problem soon becomes intractable.
First, the number of attractors grows exponentially fast with $N$;
second, larger system sizes make for increased interaction ranges
for a given node. This makes it increasingly
probable
to have winners against the odds, making it more and more difficult
to predict attractors based only on initial mass distributions.

The above algorithmic complexity goes hand in hand
with the violation of the Edwards hypothesis
as well as other specific out-of-equilibrium features of the attractors,
including a superexponential spatial decay of various correlations~\cite{us}.
This phenomenon, put forward in zero-temperature dynamics of spin chains~\cite{gds},
is reminiscent of the behaviour of a larger class of fully irreversible models,
exemplified by random sequential adsorption (RSA)~\cite{rsa}.

\section{Approximate analytical treatment}
\label{ana}

Despite the complexity referred to above, we show here that
some `one-body observables' can be predicted by an approximate analytical approach. Our techniques are based on
the inhomogeneous mean-field theory and rely
on the assumption that the statistical properties
of a node only depend on its degree~$k$~\cite{pv}.
Such ideas have been successfully applied to a wide class of problems
on complex networks (see~\cite{n3,rmp} for reviews).
The thermodynamic limit is implicitly taken; also,
the embedding graph is replaced
by an uncorrelated random graph
whose nodes have probabilities~$p_k$ to be connected to $k$ neighbors, i.e,
to have degree~$k$. In this section, we use this framework to
evaluate the survival probability of a node,
given its degree and/or initial mass.

\subsection{Survival probability of a typical node}
\label{anatypical}

We consider first the simplest observable -- the survival probability of a typical node,
irrespective of initial mass or degree.

From the reduced dynamical equations~(\ref{dadtau}),
the initial decay rate of the amplitude $a_i(\tau)$ of node $i$ is seen to be:
\beq
\omega_i=\sum_{j(i)}a_j(0)=\sum_{j(i)}y_j(0).
\label{omega}
\eeq
If node $i$ has degree $k$, the above expression is the sum of the $k$
initial masses of the neighboring nodes.

From a modelling point of view, this suggests a decimation process in continuous time,
where nodes are removed at a rate given by their degree~$k$ at time $t$.
The initial graph is entirely defined by the
probabilities $p_k$ for a node to have degree $k$.
Its subsequent evolution during the decimation process is characterized by
its time-dependent degree distribution,
i.e., by the fractions~$q_k(t)$ of initial nodes which still survive at time~$t$
and have degree $k$.
The latter quantities start from
\beq
q_k(0)=p_k
\label{qin}
\eeq
at the beginning of the process ($t=0$) and converge to
\beq
q_k(\infty)=S\,\delta_{k0}
\label{sinf}
\eeq
at the end of the process ($t\to\infty$).
Indeed, as in the original model, survivors are isolated,
so that their final degree is zero.
The amplitude $S$ is the quantity of interest,
as it represents the survival probability of a typical node.
Our goal is to determine it as a function of the probabilities~$p_k$.

The $q_k(t)$ obey the dynamical equations
\beqa
\frac{\dd q_k(t)}{\dd t}&=&-kq_k(t)
\nonumber\\
&+&\lam(t)\left((k+1)q_{k+1}(t)-kq_k(t)\right).
\label{dyns}
\eeqa
The first line corresponds to the removal of a node of degree~$k$ at constant rate $k$,
while the second line describes the dynamics of its neighbors.
The removal of one neighboring node
adds to the fraction~$q_k(t)$ at a rate proportional to $(k+1)q_{k+1}(t)$
while it depletes it at a rate proportional to $kq_k(t)$.

The time-dependent quantity $\lam(t)$
is the rate at which a random neighbor of a given node is removed at time~$t$,
which is, consistent with the above,
given by the average degree of a random neighbor of the node at time~$t$.
This rate can be evaluated as follows (see e.g.~\cite{rmp}).
The probability that a node of degree $k$ has a neighbor of degree $\ell$
at time $t$ is, for an uncorrelated network, independent of $k$, and given by
$\w q_\ell(t)=\ell q_\ell(t)/\mean{\ell(t)}$.
(The shift from $q_\ell(t)$ to $\w q_\ell(t)$ is related to the `inspection paradox'
in probability theory (see e.g.~\cite{feller}).)
The average degree of a random neighbor of the node at time~$t$ is then given by
$\sum_\ell \ell\w q_\ell(t)$, i.e.,
\beq
\lam(t)=\frac{\mean{k(t)^2}}{\mean{k(t)}}.
\label{lamdef}
\eeq

Introducing the generating series
\beq
P(z)=\sum_{k\ge0}p_kz^k,\quad Q(z,t)=\sum_{k\ge0}q_k(t)z^k,
\eeq
we see that $Q(z,t)$ obeys the partial differential equation
\beq
\frac{\dpar Q}{\dpar t}+\left(z+(z-1)\lam(t)\right)\frac{\dpar Q}{\dpar z}=0,
\label{qpar}
\eeq
with initial condition
\beq
Q(z,0)=P(z).
\eeq
Hence it is invariant along the characteristic curves defined by
\beq
\frac{\dd z}{\dd t}=z+(z-1)\lam(t).
\eeq
This differential equation can be integrated as
\beq
z_0=1+(z_t-1)\e^{-t-\Lam(t)}-\int_0^t\e^{-s-\Lam(s)}\,\dd s,
\eeq
with
\beq
\Lam(t)=\int_0^t\lambda(s)\,\dd s.
\eeq
We have therefore
\beq
Q(z_t,t)=P(z_0).
\eeq
For infinitely long times, irrespective of $z_t$,
the parameter~$z_0$ which labels the characteristic curves converges to the limit
\beq
\z=1-\int_0^\infty\e^{-t-\Lam(t)}\,\dd t,
\eeq
which we call the {\it dynamical fugacity} of the model.

We are left with the following simple expression
for the survival probability of a typical node (see~(\ref{sinf}))
\beq
S=P(\z).
\label{simp}
\eeq
Furthermore, the fugacity $\z$ can be shown to be implicitly given by
\beq
2\kave\int_\z^1\frac{\dd z}{P'(z)}=1,
\label{zimp}
\eeq
where the accent denotes a derivative.

We list a few quantitative predictions for important graphs/networks below:

\begin{itemize}

\item {\it Erd\"os-R\'enyi (ER) graph}.
This historical example of a random graph~\cite{er1,er2}
has a Poissonian degree distribution of the form
\beq
p_k=\e^{-a}\,\frac{a^k}{k!}\quad(k\ge0),
\label{erp}
\eeq
so that $\kave=a$ and $P(z)=\e^{a(z-1)}$.
We obtain
\beqa
&&\z=1-\frac{1}{a}\ln\frac{a+2}{2},
\label{zer}
\\
&&S=\frac{2}{a+2}.
\label{ser}
\eeqa

\item{\it $K$-regular graph}.
In the case of a $K$-regular graph,
where all nodes have the same degree $K\ge2$,
we have $P(z)=z^K$.
We obtain
\beqa
&&\z=\left(\frac{2}{K}\right)^{1/(K-2)},
\\
&&S=\left(\frac{2}{K}\right)^{K/(K-2)}\quad(K\ge3).
\label{sreg}
\eeqa
For $K=2$ the above results become
\beqa
&&\z=\e^{-1/2}=0.606530,
\\
&&S=\e^{-1}=0.367879.
\label{sreg2}
\eeqa

\item{\it Geometric graph}.
In the case of a geometric degree distribution with parameter $y$, i.e.,
\beq
p_k=(1-y)y^k\quad(k\ge0),
\eeq
we have $\kave=y/(1-y)$ and $P(z)=(1-y)/(1-yz)$.
We obtain
\beqa
\z&=&\frac{1}{y}-\left(\frac{(1-y)^2(2+y)}{2y^3}\right)^{1/3}
\nonumber\\
&=&1+\frac{1}{\kave}-\left(\frac{3\kave+2}{2\kave^3}\right)^{1/3},
\label{zgeo}
\\
S&=&\left(\frac{2(1-y)}{2+y}\right)^{1/3}=\left(\frac{2}{3\kave+2}\right)^{1/3}.
\label{sgeo}
\eeqa

\item{\it Barab\'asi-Albert (BA) network}.
The BA network, grown with a linear law of preferential attachment,
has a degree distribution~\cite{bas,baj}
\beq
p_k=\frac{4}{k(k+1)(k+2)}\quad(k\ge1)
\label{bapk}
\eeq
with a power-law tail with exponent $\gamma=3$.
We have $\kave=2$ and
\beq
P(z)=3-\frac{2}{z}-\frac{2(1-z)^2}{z^2}\ln(1-z).
\label{bapz}
\eeq
Solving~(\ref{zimp}) numerically, we obtain
\beq
\z=0.67016,\quad S=0.55300.
\label{zsba}
\eeq

\item{\it Generalized preferential attachment (GPA) network}.
This is a generalization of the BA network,
where the attachment probability to an existing node with degree $k$
is proportional to $k+c$~\cite{dms,krl,ggl},
with the offset $c$ representing the initial attractiveness of a node.
The degree distribution
\beq
p_k=\frac{(c+2)\Gamma(2c+3)\Gamma(k+c)}{\Gamma(c+1)\Gamma(k+2c+3)}\quad(k\ge1)
\eeq
has a power-law tail with a continuously varying exponent $\gamma=c+3$.
We have, as expected for a tree, $\kave=2$, irrespective of $c$, and
\beqa
P(z)&=&\frac{c+2}{2c+3}\,z(1-z)^{c+2}
\nonumber\\
&\times&\F(c+3,2c+3;2c+4;z),
\label{gpapz}
\eeqa
where $\F$ is the Gauss hypergeometric function.

The GPA network can be simulated efficiently
by means of a redirection algorithm~\cite{kr1,kr2}.
Every new node is attached either to a uniformly chosen earlier node
with probability $1-\nu$,
or to the ancestor of the latter node with the redirection probability
\beq
\nu=\frac{1}{c+2}.
\eeq

For $\nu\to0$ (i.e., $c\to+\infty$), we have a uniform attachment rule,
yielding $p_k=2^{-k}$ and $P(z)=z/(2-z)$,~so that
\beqa
&&\z=2-(5/2)^{1/3}=0.642791,
\label{zzero}
\\
&&S=2(2/5)^{1/3}-1=0.473612.
\label{szero}
\eeqa
For $\nu\to1$ (i.e., $c\to-1$), the model becomes singular.
The $p_k$ converge to $\delta_{k,1}$, while we still have $\kave=2$, formally.
In this singular limit we have
\beq
\z=S=\frac{3}{4}
\label{zsone}
\eeq
Finally, the BA network is recovered for $\nu=1/2$ (i.e., $c=0$).

\end{itemize}

\begin{figure}[!ht]
\begin{center}
\includegraphics[angle=-90,width=.7\linewidth]{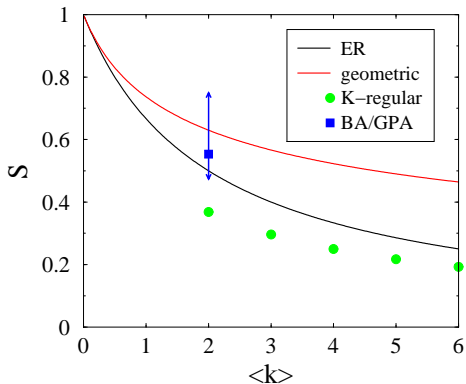}
\caption{\small
(Color online)
Survival probability $S$ of a typical node,
as predicted by our approximate analysis,
against mean degree~$\kave$ for several examples of graphs and networks.
For the ER and geometric graphs, the predictions for continuously varying $\kave$
are shown as the black (lower) and red (upper) curves.
Green filled circles correspond to $K$-regular graphs for integer $K$.
The blue square shows the prediction~(\ref{zsba}) for the BA network,
while the vertical bar with arrowheads shows the range of values of $S$ for GPA networks,
given by the bounds~(\ref{szero}) and~(\ref{zsone}).}
\label{ks}
\end{center}
\end{figure}

Figure~\ref{ks} provides a summary of the above results.
The survival probability $S$ of a typical node is plotted against the mean degree $\kave$,
for all the above.
Turning to the specific example of the GPA networks, we note that~(\ref{zzero}) and~(\ref{szero}) provide lower bounds for the dynamical fugacity~$\zeta$ and the survival probability $S$ respectively, which both increase to their upper bounds~(\ref{zsone}) as functions of
the redirection probability $\nu$.
This trend is reflected in the blue arrowheads in Figure~\ref{ks}, where the blue square corresponds to the value for the BA network.


In order to test the above for a few simple cases,
we have measured the survival probability $S$ of a typical node
in three different geometries:
the 1D chain, the 2D square lattice and the BA network.
We solved
the dynamical equations~(\ref{dadtau}) numerically,
with initial masses drawn
either from a uniform (uni) or an exponential (exp) distribution.
The measured values of $S$ are listed in Table~\ref{tabsur},
together with our approximate predictions~(\ref{sreg}),~(\ref{sreg2}) and~(\ref{zsba}).

\begin{table}[!ht]
\begin{center}
\begin{tabular}{|c|c|c|c|}
\hline
Geometry & 1D & 2D & BA\\
\hline
$S_{\rm uni} $ &$\ 0.4393\ $&$\ 0.3851\ $&$\ 0.6891\ $\\
$S_{\rm exp} $ & 0.4360 & 0.3755 & 0.6767\\
\hline
$S_{\rm pred} $ & 0.3679 & 0.2500 & 0.5530\\
\hline
\end{tabular}
\caption{Survival probability of a typical node
of the 1D chain, the 2D square lattice and the BA network.
Comparison of numerical results for uniform (uni) and exponential (exp) mass distributions
and approximate analytical results (pred).}
\label{tabsur}
\end{center}
\end{table}

Our numerical results suggest that the survival probability depends only weakly on the mass distribution,
as long as the latter is rather structureless.
While our predictions are systematically lower than the numerical
observations, they do indeed reproduce the global trends rather well.


It is worth recalling here that our analysis relies on a degree-based mean-field approach.
Such approximate techniques are expected to give poor results in one dimension,
and to perform much better on more disordered and/or more highly connected structures.
A systematic investigation of the accuracy of the present approach,
including the comparison of various lattices with identical coordination numbers
but different symmetries,
and of various random trees or networks with the same degree distribution
but different geometrical correlations,
would certainly be of great interest.
While such extensive numerical investigations
are beyond the scope of the present, largely analytical work, we hope they will be taken
up in future.

\subsection{Degree-resolved survival probability}
\label{anadegree}

The above degree-based mean-field analysis
can be extended to quantities with a richer structure.
In this section we consider the degree-resolved survival probability $S[\ell]$
of a node whose initial degree~$\ell$ is given.
The degree distribution $r_k(t)$ of this special node obeys the same dynamical
equations~(\ref{dyns}) as those of a typical node:
\beqa
\frac{\dd r_k(t)}{\dd t}&=&-kr_k(t)
\nonumber\\
&+&\lam(t)\left((k+1)r_{k+1}(t)-kr_k(t)\right),
\eeqa
with the specific initial condition $r_k(0)=\delta_{k\ell}$.
The above dynamical equations along the same lines as in Section~\ref{anatypical}.
The generating series
\beq
R(z,t)=\sum_{k\ge0}r_k(t)z^k
\eeq
obeys the partial differential equation~(\ref{qpar}),
with initial condition
\beq
R(z,0)=z^\ell.
\eeq
Using again the method of characteristics,
we get, instead of~(\ref{simp}), the following simple behavior
for the degree-resolved survival probability:
\beq
S[\ell]=\z^\ell,
\label{sell}
\eeq
where the dynamical fugacity $\z$ is given by~(\ref{zimp}).
The form of this suggests a simple physical interpretation:
the fugacity $\z$ measures the tendency of a given node to `escape' annihilation.
More quantitatively, $\z$ is the price per initial neighbor
which a node has to pay in order to survive forever.
By averaging the expression~(\ref{sell}) over the initial degree distribution $p_k$,
we recover the result~(\ref{simp}) for the survival probability $S$ of a typical node.

We now introduce the survival scale
\beq
\xi=\frac{1}{\abs{\ln\zeta}},
\eeq
so that our key result~(\ref{sell}) reads
\beq
S[\ell]=\exp(-\ell/\xi).
\eeq
This representation makes it clear that the survival scale~$\xi$ corresponds to the degree
of the most connected survivors.

We use this to probe the survival statistics of highly connected graphs.
Here,
the survival scale $\xi$ is large as a consequence of a large mean degree $\kave$;
correspondingly, there is a decay in the survival probability $S$ of a typical node,
as shown in the two examples below:

\begin{itemize}

\item {\it ER graph}.
Here, $\kave=a$, and so~(\ref{zer}) and~(\ref{ser}) yield
\beq
\xi\approx\frad{\kave}{\ln\frad{\kave}{2}},\quad
S\approx\frac{2}{\kave}.
\label{xic}
\eeq
The survival scale $\xi$ grows almost linearly with $\kave$,
while the survival probability $S$ falls off as $1/\kave$.

\item {\it Geometric graph}.
Equations~(\ref{zgeo}) and~(\ref{sgeo}) yield
\beq
\xi\approx\left(\frac{2\kave^2}{3}\right)^{1/3},\quad
S\approx\left(\frac{2}{3\kave}\right)^{1/3}.
\eeq
Both the growth of the survival scale and the decay of the survival probability
are slower than in the ER case.
The survival scale diverges sublinearly with~$\kave$, with an exponent $2/3$, while
the survival probability decays with an exponent~$1/3$.

\end{itemize}

Figure~\ref{ks} shows that these differing trends for the ER and geometric graphs
are already evident even for low connectivity~$\kave$.
This is because of the interesting
circumstance that the behavior of the degree distribution~$p_k$
for relatively small degrees ($1\ll k\ll\kave$) determines
both the growth of the survival scale~$\xi$
as well as the decay of the survival probability $S$ in highly connected graphs.

Assuming this regime is described by a scaling form
\beq
p_k\approx\frac{C\,k^{\beta-1}}{\kave^\beta}
\eeq
governed by an exponent $\beta>0$,
we obtain after some algebra
\beqa
\xi&\approx&A\;\kave^{(\beta+1)/(\beta+2)},
\label{xib}
\\
S&\approx&B\;\kave^{-\beta/(\beta+2)},
\label{sb}
\eeqa
with
\beqa
A&\approx&\left(\frac{2}{\beta(\beta+2)\,C\,\Gamma(\beta)}\right)^{1/(\beta+2)},
\\
B&\approx&\left(\frac{2^\beta\,(C\,\Gamma(\beta))^2}{(\beta(\beta+2))^\beta}\right)^{1/(\beta+2)}.
\eeqa
The exponents $(\beta+1)/(\beta+2)$ and $\beta/(\beta+2)$
which enter the power laws~(\ref{xib}),~(\ref{sb})
are always smaller than one (the value corresponding to $\beta\to\infty$),
when we find $\xi\sim\kave$ and $S\sim1/\kave$.
This is e.g.~the case for the ER graph (see~(\ref{xic})).

\begin{figure}[!ht]
\begin{center}
\includegraphics[angle=-90,width=.7\linewidth]{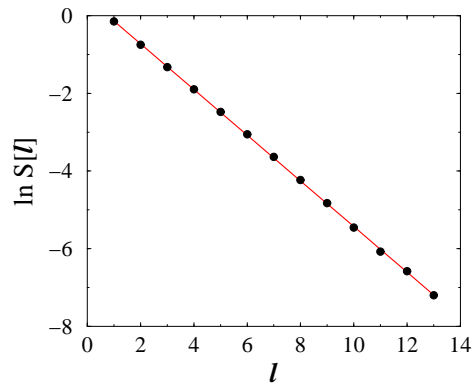}
\caption{\small
(Color online)
Logarithmic plot of the measured degree-resolved
survival probability $S[\ell]$ against degree $\ell$ for the BA network.
Straight line: least-squares fit of all the data points with slope $-0.588$.}
\label{bak}
\end{center}
\end{figure}

In order to test our key prediction~(\ref{sell}),
the dependence of the survival probability $S[\ell]$
of a node of the BA network on its degree $\ell$
was computed numerically for an exponential distribution of initial masses.
Figure~\ref{bak} shows a logarithmic plot of $S[\ell]$ against degree~$\ell$.
We find excellent qualitative agreement with our predictions of exponential dependence,
although the measured slope, corresponding to
$1/\xi_{\rm obs}\approx0.59$ is larger than our analytical prediction of
$1/\xi_{\rm pred}=\abs{\ln\zeta}=0.4002$ (see~(\ref{zsba})).

\subsection{Mass-resolved survival probability}
\label{anamass}

Here, we extend our analysis to the mass-resolved survival probability
of a node whose initial mass $y$ is given.
Since this will only enter through the reduced mass $\a$,
defined as the dimensionless ratio
\beq
\a=\frac{y}{\mean{y}},
\label{adef}
\eeq
we denote the mass-resolved survival probability by $S_\a$.

Along the lines of Section~\ref{anatypical},
we derive the following dynamical equations
for the degree distribution $r_k(t)$ of the special node:
\beqa
\frac{\dd r_k(t)}{\dd t}&=&-kr_k(t)
\nonumber\\
&+&\lam_\a(t)\left((k+1)r_{k+1}(t)-kr_k(t)\right),
\eeqa
with initial condition $r_k(0)=p_k$ (see~(\ref{qin})).
To compute the rate $\lam_\a(t)$ at which a random neighbor of the special node is
removed, we argue as follows.
This rate is a product
of the probability of finding a random neighbor of degree $\ell$,
given by $\w q_\ell(t)=\ell q_\ell(t)/\mean{\ell(t)}$,
and the rate of removal of this node.
The latter is nothing but its effective degree:
while the initial masses of its $\ell-1$ typical neighbors
can be taken to be $\mean{y}$, the special node has an
initial mass of $y=\a\mean{y}$, so that the effective degree of this node is $\ell-1+\a$.
The average degree of a random neighbor of the special node at time $t$
is then clearly given by $\sum_\ell (\ell-1+\a)\w q_\ell(t)$, i.e.,
\beq
\lam_\a(t)=\lam(t)-1+\a.
\label{lama}
\eeq

After some algebra, the mass-resolved survival probability reduces to:
\beq
S_\a=P(\z_\a),
\label{smass}
\eeq
where the mass-dependent fugacity $\z_\a$ is
\beq
\z_\a=1-\int_0^\infty\e^{-\a t-\Lam(t)}\,\dd t.
\label{zadef}
\eeq
This formula can be recast as
\beq
\z_\a=1-\int_\z^1\left(1-2\kave\int_z^1\frac{\dd y}{P'(y)}\right)^{(\a-1)/2}\dd z,
\label{zmass}
\eeq
where $\z$ is the fugacity given by~(\ref{zimp}).
We have, consistently, $\z_\a=\z$ for $\a=1$.

The mass-resolved survival probability $S_\a$
is an increasing function of the reduced mass $\a$.
We compute this explicitly for the classes of random graphs and networks considered in Section~\ref{anatypical}.

\begin{itemize}

\item {\it ER graph}.
In this case, we have
\beq
S_\a=\e^{a(\z_\a-1)},
\eeq
with
\beq
\z_\a=1-\frac{1}{a}
\int_0^a\left(\frac{a(x+2)}{(a+2)x}\right)^{(1-\a)/2}\frac{\dd x}{x+2}.
\eeq

\item{\it $K$-regular graph}.
In this case, we have
\beq
S_\a=\z_\a^K,
\eeq
with
\beq
\z_\a=1-\frac{1}{2}\int_0^1(2x^2+K(1-x^2))^{-(K-1)/(K-2)}x^\a\dd x
\eeq
in the generic case ($K\ge3$), whereas
\beq
\z_\a=1-\int_0^1\e^{(x^2-1)/2}\,x^\a\dd x
\eeq
for $K=2$.

\item{\it Geometric graph}.
In this case, we have
\beq
S_\a=\frac{1-y}{1-y\z_\a},
\eeq
with
\beq
\z_\a=1-\frac{1}{2}\int_0^1
\left(\frac{2(1-y)}{2+(1-3x)y}\right)^{2/3}x^{(\a-1)/2}\,\dd x.
\eeq

\item{\it BA and GPA networks}.
In these cases, numerical values of $\z$ and of $\z_\a$
can be extracted from~(\ref{zimp}) and~(\ref{zmass}),
using the expressions~(\ref{bapz}),~(\ref{gpapz}) of the generating series $P(z)$.

\end{itemize}

\begin{figure}[!ht]
\begin{center}
\includegraphics[angle=-90,width=.7\linewidth]{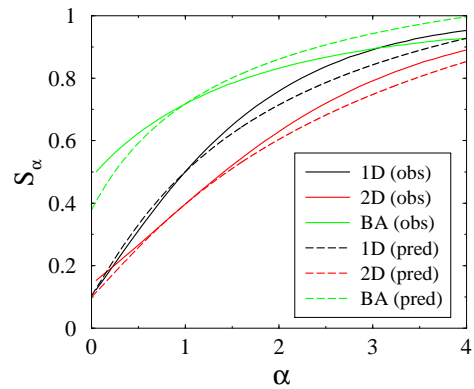}
\caption{\small
(Color online)
Mass-resolved survival probability $S_\a$ against reduced mass $\a$.
Full curves: numerical results.
Corresponding dashed curves: analytical predictions.
Top to bottom near $\a=1$: BA network, 1D chain, 2D square lattice.}
\label{y}
\end{center}
\end{figure}

The dependence of the survival probability of a node on its initial mass
was computed numerically
for the 1D chain, the 2D square lattice and the BA network,
with an exponential mass distribution.
Figure~\ref{y} shows plots of the measured values of $S_\a$ against $\a$.
The dashed curves show the prediction~(\ref{smass}),
rescaled so as to agree with the numerics for $\a=1$.
In the three geometries considered,
the analytical prediction reproduces the overall mass
dependence of $S_\a$ reasonably well.
The observed dependence is slightly more pronounced than predicted
on the 1D and 2D lattices, while the opposite holds for the BA network.

Last but by no means least, there is a striking manifestation of (super-)universality in the regime of large reduced
masses, when~(\ref{zadef}) simplifies to
\beq
\z_\a\approx1-\frac{1}{\a}.
\label{zauni}
\eeq
The mass-resolved survival probability then
goes to unity according to the simple universal law
\beq
S_\a\approx1-\frac{\kave}{\a}.
\label{slarg}
\eeq

This key result in the large-mass limit is one of the strongest results in this paper: all details
of the structure and embeddings of networks disappear from the survival probability of a node,
leaving only a simple dependence on its mass and the mean connectivity $\kave$.

\subsection{Degree and mass-resolved survival probability}
\label{anadm}

Finally, our analysis
can be generalized to the full degree and mass-resolved survival probability $S_\a[\ell]$
of a node whose initial degree $\ell$ and reduced mass $\a$ are given.

The degree distribution of this special node obeys
\beqa
\frac{\dd r_k(t)}{\dd t}&=&-kr_k(t)
\nonumber\\
&+&\lam_\a(t)\left((k+1)r_{k+1}(t)-kr_k(t)\right),
\eeqa
where the rate $\lam_\a(t)$ is given by~(\ref{lama}),
and with the specific initial condition $r_k(0)=\delta_{k\ell}$.
After some algebra along the lines of the previous sections, we find:
\beq
S_\a[\ell]=\z_\a^\ell.
\label{sfull}
\eeq

This last result encompasses all the previous ones,
including the expression~(\ref{sell})
for the degree-resolved survival probability $S[\ell]$
and the expression~(\ref{smass})
for the mass-resolved survival probability $S_\a$.

The detailed numerical evaluation of degree and mass-resolved data on
networks is deferred to future work, although we expect the levels of agreement to
to be similar to those obtained in Sections~\ref{anadegree} and~\ref{anamass}.
Our point of emphasis here is the simplicity of~(\ref{sfull}), which
shows that the dynamical fugacity is absolutely the right variable to highlight
an intrinsic universality in this problem. In showing that the result~(\ref{simp}) is resolvable
into components of degree and mass, it also points the way towards a deeper understanding
of the `independence' of these parameters in the survival of a node.

Finally, and no less importantly,
the result~(\ref{sfull}) again manifests (super-)universality in the regime of large reduced
masses ($\a\gg1$).
As a consequence of~(\ref{zauni}),
the degree and mass-resolved survival probability
goes to unity according to the simple law
\beq
S_\a[\ell]\approx1-\frac{\ell}{\a}.
\label{sfuller}
\eeq

The beauty of this result (as well as its analogue~(\ref{slarg})) lies in the fact that it is exactly
what one might expect intuitively; it suggests that the probability ($1 - S_\a[\ell]$) 
that a node of mass $\a$ and degree $\ell$ might {\it not} survive,  is directly proportional
to its degree and inversely proportional to its mass. In everyday terms, the lighter the node,
and the more well-connected it is, the more it is likely to disappear. 
The emergence of such startling, intuitive simplicity in an extremely complex system is a testament to an underlying elegance
in this model.

\section{Discussion}

The problem of finding even an approximate analytical solution to a model which
contains multiple interactions is very challenging.
In the context of predator-prey models, the (mean-field) Lotka-Volterra dynamical system and the full Volterra chain
are among the rare examples which are integrable.
Most other nonlinear dynamical models with competing interactions
are not, and are quite simply intractable analytically.

The model inspired by black holes~\cite{us},
on which this paper is based, shows how competition between local
and global interactions can give rise to non-trivial survivor patterns,
and to the phenomenon referred to as `winning against the odds'.
That is, a given mass can win out
against more massive competitors in its immediate neighborhood provided that
they in turn are `consumed' by ever-more distant neighbors.
When it was found numerically~\cite{c1,c3,c2} that such survivor distributions
seemed to exhibit somewhat surprising features of universality,
it was natural to ask the question:
could one find the reasons for such behaviour,
in the sense of characterizing these distributions at least approximately
from an analytic point of view?
An additional motivation was found in the work of the Barab\'asi group~\cite{wsb}
on citation networks, where the authors put forward a universal scaling form
for the `survival' of a paper in terms of its citation history.

The black-hole model as defined in its original cosmological context~\cite{bh1,bh2}
had an explicit time dependence due to the presence of a ubiquitous `fluid',
as well as a threshold below which even isolated particles did not survive.
Neither of these attributes was necessary for the behaviour of most interest to us,
namely the multiplicity of attractors (which in our case involve survivor distributions),
and their non-trivial dependence on the initial mass profile,
as a result of multiple interactions.
One of our major achievements in this paper has been the construction of
a much simpler model (without the unnecessary complications referred to above),
which still retains its most interesting features from the point of view of statistical physics.

Once derived and established,
this simple model was the basis of our investigations of
universality in survivor distributions.
The exact characterisation of attractors
as a function of the initial data becoming rapidly impossible,
we were led to think of approximate analytical techniques.
Our choice of the inhomogeneous (or degree-based) mean-field theory
was motivated by our emphasis on random graphs and networks
in earlier numerical work~\cite{c1,c3,c2}.
This approach was embodied in an effective decimation process.
Some of the analytical results so obtained were robustly universal,
including the exponential fall-off~(\ref{sell})
of the survival probability of a node with its degree,
or the asymptotic behaviours~(\ref{slarg}) and~(\ref{sfuller})
 in the large-mass regime.

Our approach led us to introduce the associated concept of a dynamical fugacity,
key to unlocking the reason behind the manifestation of universality
in diverse survivor distributions.
Physically, this signifies the tendency of a typical node in a network
to escape annihilation, which we illustrate via a simple argument.
Every time an agent encounters other, potentially predatory, agents it pays a
price in terms of its survival probability:
as the probability of each such encounter is independent of the others,
the `cost' to the total probability is multiplicative in terms of
the number of predators encountered.
The dynamical fugacity is then nothing but the cost function per encounter
(i.e., per neighbor),
so that the survival probability of the original agent
depends exponentially on the number of its competitors.
In a statistical sense, this depends only on the degree distribution
of the chosen network, leading to the emergence of a universal survival
probability for a given class of networks.
The asymptotic survival probability of very heavy nodes
becomes `super-universal', in the sense of losing all dependence
on different geometrical embeddings. Its form also has an appealing simplicity,
as the complement of the ratio (degree/mass) of a given node; the heavier the node
and the more isolated it is, the longer it is likely to survive.

In conclusion, we have used inhomogeneous mean-field theory to formulate and solve
analytically an intractable problem with multiple interactions.
While our analytical solutions are clearly not exact (due to the
technical limitations of mean-field theory) they are nevertheless
the only way to date of understanding the behaviour of the exact system.
In
particular, and importantly,
our present analysis strongly reinforces the universality that has indeed been
observed in earlier numerical simulations of this problem~\cite{c1,c3,c2}.
That such universal features emerge in a highly complex many-body problem
with competing predatory interactions, is nothing short of remarkable.

\begin{acknowledgments}

It is a pleasure to thank Olivier Babelon and Alfred Ramani for interesting discussions.

\end{acknowledgments}

\appendix

\section{Numbers of attractors and complexity in one dimension}
\label{app}

In this appendix we investigate the attractor statistics of the one-dimensional
problem by means of the transfer-matrix formalism.
We describe an attractor as a sequence of binary variables or spins:
\beq
\s_n=\left\{\matrix{
1&\mbox{if $i$ is a survivor},\hfill\cr
0&\mbox{if $i$ is a non-survivor}.\hfill\cr
}\right.
\eeq
From a static viewpoint, attractors are defined as patterns obeying the
constraints listed in Section~\ref{model}.
They can therefore be identified with sequences which avoid the patterns 11 and 000.
The last two symbols of such a sequence may therefore be 00, 01 or 10.
The numbers $M_N^{00}$, $M_N^{01}$ and $M_N^{10}$ of attractors of length $N$
of each kind obey the recursion
\beq
\pmatrix{M_{N+1}^{00}\cr M_{N+1}^{01}\cr M_{N+1}^{10}}
=\T
\pmatrix{M_N^{00}\cr M_N^{01}\cr M_N^{10}},
\eeq
where the transfer matrix $\T$ reads
\beq
\T=\pmatrix{0&0&1\cr 1&0&1\cr 0&1&0}.
\eeq
Its characteristic polynomial is $P(x)=x^3-x-1$,
and so the Cayley-Hamilton theorem implies the recursion
\beq
\T^N=\T^{N-2}+\T^{N-3}.
\label{ch}
\eeq
Hence all the numbers $M_N$ grow exponentially with~$N$,
in agreement with~(\ref{sdef}).
The complexity reads
\beq
\Sigma=\ln x_0=0.281199,
\label{sres}
\eeq
with $x_0=1.324717$ being the largest eigenvalue of $\T$,
i.e., the largest root of $P(x)$.

The total numbers of attractors $M_N^\c=M_N^{00}+M_N^{01}+M_N^{10}$ on chains
and $M_N^\r$ on rings of $N$ nodes obey recursions derived from~(\ref{ch}), i.e.,
\beq
M_N=M_{N-2}+M_{N-3},
\eeq
with two different sets of initial conditions.
The sequences $M_N^\r$ and $M_N^\c$
are listed in the OEIS~\cite{OEIS}, respectively as entries A001608 and A000931,
together with many combinatorial interpretations and references.
The first few terms are listed in Table~\ref{numbers}.

\begin{table}[!ht]
\begin{center}
\begin{tabular}{|c||c|c|c|c|c|c|c|c|c|c|c|c|}
\hline
$N$&$\ 1\ $&$\ 2\ $&$\ 3\ $&$\ 4\ $&$\ 5\ $&$\ 6\ $&$\ 7\ $&$\ 8\ $&$\ 9\ $&$\
10\ $&$\ 11\ $&$\ 12\ $\\
\hline
$M_N^\r$ & 0 & 2 & 3 & 2 & 5 & 5 & 7 & 10 & 12 & 17 & 22 & 29\\
\hline
$M_N^\c$ & 1 & 2 & 2 & 3 & 4 & 5 & 7 & 9 & 12 & 16 & 21 & 28\\
\hline
\end{tabular}
\caption{Numbers of attractors $M_N^\r$ on rings and $M_N^\c$ on chains
of $N$ nodes, for $N$ up to 12.}
\label{numbers}
\end{center}
\end{table}

\begin{figure}[!ht]
\begin{center}
\includegraphics[angle=-90,width=.7\linewidth]{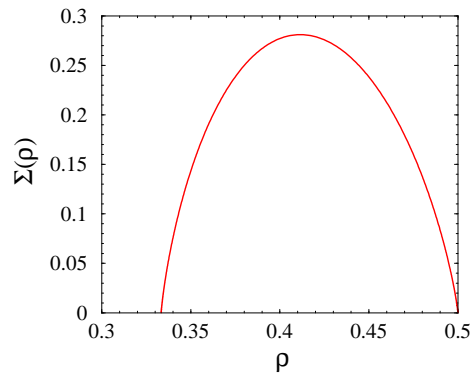}
\caption{\small
(Color online)
Density-dependent static complexity $\Sigma(\rho)$ against survivor density
$\rho$ in the allowed range ($1/3\le\rho\le 1/2$).}
\label{sent}
\end{center}
\end{figure}

The transfer-matrix approach can be generalized
in order to determine the density-dependent static complexity $\Sigma(\rho)$,
characterizing the exponential growth of the number of attractors
with a fixed density $\rho$ of survivors.
Introducing a static fugacity $z$ conjugate to the number of survivors,
the transfer matrix becomes
\beq
\T(z)=\pmatrix{0&0&1\cr z&0&z\cr 0&1&0},
\eeq
whose characteristic polynomial is $P(z,x)=x^3-zx-z$.
The reader is referred to~\cite{gds,glrev} for details.
We obtain after some algebra
\beq
\Sigma(\rho)=-(1-2\rho)\ln\frac{1-2\rho}{\rho}-(3\rho-1)\ln\frac{3\rho-1}{\rho}.
\label{srho}
\eeq
Figure~\ref{sent} shows a plot of this quantity
against survivor density in the allowed range ($1/3\le\rho\le 1/2$).
The expression~(\ref{srho}) reaches a maximum equal to $\Sigma$ (see~(\ref{sres}))
when $\rho$ equals the mean static density of survivors:
\beq
\rho_0=\frac{x_0+1}{2x_0+3}=0.411495.
\label{rhozero}
\eeq

\bibliography{revised.bib}

\end{document}